\documentclass[runningheads]{llncs}
\usepackage[utf8]{inputenc}
\usepackage[T1]{fontenc}
\usepackage[pdftex]{graphicx}
\usepackage{microtype}
\usepackage{csquotes}
\usepackage{gensymb}
\usepackage{todonotes}
\usepackage{amsmath}
\usepackage{amssymb}
\usepackage[toc,page]{appendix}
\usepackage{booktabs}
\newcommand{\otoprule}{\midrule[\heavyrulewidth]}
\usepackage{pifont}% http://ctan.org/pkg/pifont
\newcommand{\cmark}{\ding{51}}%
\newcommand{\xmark}{\ding{55}}%

\makeatletter
\def\@fnsymbol#1{\ensuremath{\ifcase#1\or *\or \dagger\or \ddagger\or
   \mathsection\or \mathparagraph\or \|\or **\or \dagger\dagger
   \or \ddagger\ddagger \else\@ctrerr\fi}}
\makeatother

\begin{document}
%

%\title{Divide and Conquer: A Hierarchical Deep Neural Network for Anatomical Shapes}
\title{Is a PET all you need? A multi-modal study for Alzheimer's disease using 3D CNNs }
\titlerunning{Is a PET all you need?}
\author{Marla Narazani$^{1,}$\thanks{Corresponding author: marla.narazani@tum.de}$^{,}$\thanks{These authors contributed equally to this work.}, Ignacio Sarasua$^{1,2,\dag}$, Sebastian Pölsterl$^{2}$, Aldana Lizarraga$^{1}$, Igor Yakushev$^{1}$, Christian Wachinger$^{1,2}$}
% index{Narazani, Marla} 
% index{Sarasua, Ignacio} 
% index{P{\"{o}}lsterl,Sebastian}
% index{Lizarraga, Aldana} 
% index{Yakushev, Igor} 
% index{Wachinger, Christian}
\authorrunning{M. Narazani, I. Sarasua, et al.}
%\institute{Under Review at MICCAI 2022}
\institute{%
$^{1}$Technical University of Munich, School of Medicine \\
$^{2}$Lab for Artificial Intelligence in Medical Imaging (AI-Med), KJP, LMU Klinikum 
}
\maketitle              % typeset the header of the contribution

\begin{abstract}
Alzheimer's Disease (AD) is the most common form of dementia
and often difficult to diagnose due to the multifactorial etiology of dementia.
Recent works on neuroimaging-based computer-aided diagnosis 
with deep neural networks (DNNs) showed that fusing
structural magnetic resonance images (sMRI) and
fluorodeoxyglucose positron emission tomography (FDG-PET)
leads to improved accuracy in a study population of healthy
controls and subjects with AD.
However, this result conflicts with the established clinical knowledge that FDG-PET better captures AD-specific pathologies
than sMRI.
Therefore,
we propose a framework for the systematic evaluation of multi-modal DNNs
and critically re-evaluate single- and multi-modal
DNNs based on FDG-PET and sMRI for binary healthy vs.\ AD, and three-way healthy/mild cognitive impairment/AD classification.
Our experiments demonstrate that a single-modality network
using FDG-PET performs better than MRI (accuracy 0.91 vs 0.87) and does not show improvement when combined.
This conforms with the established clinical knowledge on
AD biomarkers, but raises questions about the true
benefit of multi-modal DNNs.
We argue that future work on multi-modal fusion should
systematically assess the contribution of individual
modalities following our proposed evaluation framework.
Finally, we encourage the community to go beyond
healthy vs.\ AD classification and focus on differential
diagnosis of dementia, where fusing multi-modal image
information conforms with a clinical need.
\end{abstract}

\section{Introduction}

With life expectancies rising globally,
dementia is becoming a growing concern
for individuals and society.
Dementia is characterized by a progressive cognitive impairment
that eventually requires
individuals to be completely dependent upon caregivers.
While this process cannot be reversed, 
recent efforts have focused on diagnosing subjects at an early stage to improve disease management~\cite{Borson2013}.
A particular focus has been on Alzheimer's Disease (AD), given that it is the most common form of dementia and benefits from large data-sharing initiatives~\cite{Livingston2017}.
To date, a wide range of diagnostic tools are available
for diagnosing AD: magnetic resonance imaging (MRI),
positron emission tomography (PET),
cerebrospinal fluid (CSF), demographics,
cognitive tests, and genetic alterations~\cite{Aisen2017}.
Structural MRI (sMRI) captures regional atrophy of the brain, whereas FDG-PET measures the brain's glucose metabolism.
FDG-PET plays a major role in the clinical diagnosis of AD.
It can detect functional brain changes in AD early in the disease progression and can help to differentiate AD from other causes of dementia such as frontotemporal and Lewy body dementia~\cite{marcus2014brain}.
In the memory clinic,
MRI and FDG-PET are among the most common neuroimaging methods used~\cite{Teipel2017},
where FDG-PET is considered to have a higher
diagnostic and prognostic accuracy~\cite{bloudek2011review,Frisoni2013}.

% In fact, it has been shown to have superior diagnostic accuracy compared to other diagnostic methods such as clinical guidelines, MRI, CT, SPECT, and biomarkers \cite{bloudek2011review}. Several studies comparing FDG and sMRI on the basis of diagnostic and prognostic accuracy have found FDG to provide better results than MRI \cite{mosconi2006hypometabolism,de2001hippocampal}, particularly for early-onset subjects \cite{matsunari2007comparison}.
% Though there is no consensus regarding the best diagnostic method in a memory clinic setting, CT scans, sMRI and FDG-PET are among the most common neuroimaging methods used  \cite{Teipel2017}. However, given that FDG-PET changes might precede MRI changes \cite{jack2010hypothetical} and its better prediction accuracy, it has been gained significant popularity in clinical settings \cite{burge2018empfehlungen,ossenkoppele2013impact,marcus2014brain}.

Recently, studies on deep learning (DL) techniques have emerged
that showed that distinguishing healthy controls from
AD subjects becomes more accurate when
learning from MRI \emph{and} FDG-PET, rather than
a single modality~\cite{song2021effective,zhang2011multimodal,zhou2019effective}.
However, this scenario is very different from that in
a memory clinic. In the clinic, the main objective is differential
diagnosis to determine the type of dementia, whereas
studies on DL merely considered a single type of dementia,
namely AD~\cite{song2021effective,zhang2011multimodal,zhou2019effective}.
When considering that both modalities assess neural degeneration, but
AD-specific changes are better captured by FDG-PET
than MRI~\cite{bloudek2011review,Frisoni2013},
it seems surprising why combining MRI and FDG-PET
with DL would be beneficial when AD is the only form of dementia that is being studied.

In this work, we critically re-evaluate single- and multi-modal DL models
based on FDG-PET and structural MRI for
classifying healthy vs. AD subjects.
We study three different modes of multimodal fusion:
early, middle, and late fusion.
We evaluate each to investigate whether it truly
benefits from multi-modal data by performing
ablation studies for which MRI and FDG-PET images
are paired randomly.
Contrary to previous work, our experiments show
that FDG-PET alone is sufficient for AD diagnosis,
which conforms with established clinical knowledge about biomarkers in AD.
We argue that future work on multi-modal fusion should
follow our proposed evaluation framework
to systematically assess the contribution of individual modalities.
\subsubsection{Related work.}
% Earlier methods focused on using linear classifiers such as support vector machines (SVM) based on hand-crafted features \textcolor{red}{TODO}. However, given that such methods require expert knowledge for feature extraction and are time consuming, they have now been replaced by more sophisticated deep learning algorithms. Such methods eliminate the need for manually extracting features and can be trained from raw or less pre-processed data. Especially Convolutional Neural Networks (CNN), have been found to outperform existing machine learning methods in image classification tasks \cite{lecun2015deep}.
Most DL models for AD prediction are single-modal
(see \cite{ebrahimighahnavieh2020deep} for an overview).
In \cite{farooq2017deep}, the authors propose a 2D convolutional neural network (CNN)
using slices of sMRI volumes.
However, recent work has shifted towards 3D CNN architectures for analyzing sMRI \cite{basaia2019automated,esmaeilzadeh2018end,hosseini2016alzheimer,korolev2017residual,li2017alzheimer,payan2015predicting}.
A sparse autoencoder is combined with a CNN in~\cite{payan2015predicting}.
Korolev et al.~\cite{korolev2017residual} compare a 3D-VGG and 3D-Resnet architecture. Both \cite{basaia2019automated} and \cite{esmaeilzadeh2018end} use a 3D CNN for whole brain MRIs. 
Regarding work related to FDG-PET,
a 2D CNN has been used in~\cite{ding2019deep,liu2018classification},
and a 3D CNN in~\cite{yee2020quantifying}.
Finally, several works combined sMRI and FDG-PET \cite{song2021effective,zhang2011multimodal,zhou2019effective}.
In \cite{song2021effective}, the authors propose an early fusion approach by overlaying gray matter (GM) tissues from MRI with the FDG-PET scans
and evaluate the effectiveness of their fusion strategy using a 3D CNN.
% and evaluate using a 3D CNN and a 3D Multi-scale CNN to evaluate the effectiveness of their image fusion.
In \cite{zhou2019effective}, a three-stage framework based on middle and late fusion using MRI, FDG-PET, and single nucleotide polymorphisms
is proposed.
The authors of \cite{feng2019deep}
combine a 3D CNN and LSTM.
Finally, in \cite{huang2019diagnosis},
an early and a late fusion approach are presented based on a 3D-VGG.
The works on multi-modal fusion
unanimously concluded that fusing sMRI
and FDG-PET improves prediction
accuracy over using a single modality,
which conflicts with the
established clinical knowledge that FDG-PET better captures AD-specific pathologies
than sMRI~\cite{bloudek2011review,Frisoni2013}.

%%%%%%%%%%%%%%%%%%%%%%%%%
% Methods
%%%%%%%%%%%%%%%%%%%%%%%%%
\section{Methods}\label{sec:method}
To determine the contribution of each modality in a multi-modal DNN,
we propose a systematic evaluation framework.
First, we consider
each modality in isolation by using a single branch 3D CNN.
Next, we consider the joint contribution of
multiple modalities using a 3D CNN with
either early, late, or middle fusion (see Fig.~\ref{fig:overview}).
% as described in \cite{boulahia2021early}.
To assess whether multi-modal inputs are truly helpful, we perform ablation experiments where
MRI and FDG-PET images are paired randomly.
This allows us to quantify to importance
of each modality.

% It is important to note that the same individual processing should be applied to both modalities for all combinations to avoid bias introduced by different data processing.

% Additionally, \todo{keep network architecture more similar to avoid bias, simple network for single modality vs deep network for multi-modal.}

\subsection{CNN Architecture}\label{sec:backbone}
We use a 3D ResNet as the base architecture for all models (more details in supplemental Fig.~\ref{fig:network_architecture}). It comprises 12 convolutional layers with kernel size $3^3$ in total. We use four residual learning blocks consisting of two convolutional layers followed by batch normalization (BN)~\cite{Ioffe2015} and rectified linear unit (ReLU) activation. We half the spatial resolution of feature maps in the last three residual blocks by using a stride of 2. %It is followed by 
%one residual block without downsampling,
% three more residual blocks with downsampling.
Finally, we perform global average pooling across the
spatial dimensions of the feature maps and use two linear
layers to output a log-probability. We use dropout in each residual block to reduce overfitting.

\subsection{Fusion Strategies}
\begin{figure}[t]
	\centering
	\includegraphics[width=\linewidth]{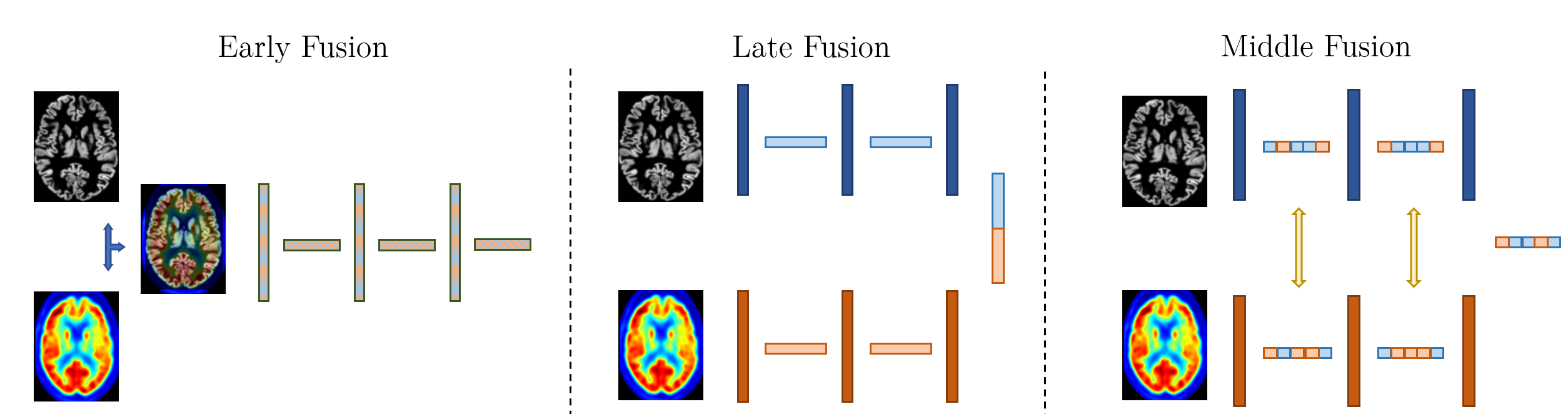}
	\caption{Overview of the three fusion strategies. Early fusion combines the MRI and PET inputs in a single volume. Late fusion concatenates the latent representations coming from each independent network. Middle fusion exchanges channels of the intermediate feature maps along the network. }
	\label{fig:overview}
\end{figure}

We consider three strategies for fusing multi-modal data: 
%Many methods have been proposed for fusing multi-modal data. In this work we want to focus on three strategies depending when the different volumes are fused with one another: 
early, late, and middle fusion (see Fig.~\ref{fig:overview}).
All three strategies follow the base
CNN architecture described above.
Next, we describe the fusion
strategies in detail.

\textbf{Early Fusion.}\label{sec:early_fusion}
In early fusion, raw modalities are combined
directly before being passed to the network.
Here, we follow the strategy proposed in \cite{song2021effective}:
gray matter maps are obtained via Voxel-Based Morphometry (VBM) and used to mask the FDG-PET intensities.
In the resulting volume, the intensities of
the FDG-PET are effectively weighted by the MRI intensities.
% Both images are previously processed and co-registered.
The network is a single branch network that
receives the combined MRI-FDG-PET volume as input.

\textbf{Late Fusion.}
Late fusion is the most straight-forward
approach to fuse multi-modal data.
Rather than fusing the images, it
fuses the latent representations of
two separate networks.
Here, we train two independent 3D ResNet branches,
one for MRI and one for FDG-PET.
The features obtained from each branch after global average pooling are then concatenated and passed through a Multi-layer perceptron (MLP) [128, 64, number of classes] to obtain a log-probability
that accounts for both sources of information.

\textbf{Middle Fusion.}
While early and late fusion are common in multi-modal analysis,
we also explore an approach that
fuses intermediate representations of
modality-specific networks~\cite{wang2020deep}.
In this approach, modality-specific information are fused by dynamically exchanging feature maps between sub-networks of different modalities.
This bi-directional exchange of information is self-guided by considering individual channel importance, which is measured by the magnitude of the BN scaling factor.
This process is carried out under the $\ell_{1}$ regularization
to penalize exchanging all channels. 
To further encourage sharing of information,
convolutional filter weights are shared across modalities.
Note that BN layers are not shared in order to determine the channel importance for each individual modality.
% which enforces a penalty and prunes out filters meeting a sparsity criteria.
To the best of our knowledge, channel exchange has not been applied for multi-modal fusion
for AD prediction before.

\subsection{Evaluation Scheme}\label{sec:eval_scheme}
Our main objective is to rigorously evaluate whether MRI is truly relevant for diagnosing AD when FDG-PET is available too.
For all of our experiments, we train the models using FDG-PET and MRI data
from the same patient. During inference
we define three different experiments based on the
input data: (i) correct, (ii) random PET, and (iii) random MRI.
We use balanced accuracy (BACC) to assess the
predictive performance of models, because
it is insensitive to the relative frequency
of class labels~\cite{Brodersen2010}.

% Our main objective is to rigorously evaluate whether sMRI is truly relevant for diagnosing AD when FDG-PET is available.
% We propose three experiments for each of the fusion methods described above.
% In all experiments, 
% During inference, we define three different types of input:
% remains associated to a specific patient,
% but the MRI is exchanged.

\noindent
\textbf{Correct.}
This strategy follows the standard training and
testing scheme.
For each fusion strategy, we test
the networks based on FDG-PET and MRI scans from the same patient.
% (and registered between them). 
If both modalities would be relevant for AD diagnosis,
we would expect this scenario to yield the highest
predictive performance.
It serves as a baseline for the remaining experiments.

\noindent
\textbf{Random MRI.}
In this experiment, we pair a patient's true
FDG-PET image and diagnosis with
an MRI of a randomly selected patient.
%We replicate the same experiment previously described but this time the MRI scan is from a different patient as the FDG-PET. 
% The model is trained according to the PET's diagnosis.
If both modalities would be relevant for the final decision, we would expect a significant drop in performance with respect to the \emph{Correct MRI} experiment.
Otherwise, if performance remains similar,
the contribution of patient-specific anatomy,
as captured by the MRI,
the MRI adds little additional information
that is not available from the FDG-PET.
% to the overall prediction is negligibly small.

\noindent
\textbf{Random PET.}
This experiment is similar
to the previous experiment, but
this time we pair the correct MRI
and diagnosis with a randomly selected FDG-PET from another patient. 
%If both modalities were as relevant for diagnosis, we would expect the performance
%of this experiment to be similar to the previous one, and much lower than when 
%testing in the correct pair of scans.
The conclusions we can derive from this
experiment are the same as in the previous
experiment, but focus on assessing the
contribution of FDG-PET.

%%%%%%%%%%%%%%%%%%%%%%%%%
% Experiments
%%%%%%%%%%%%%%%%%%%%%%%%%
% \section{Experiments}
\subsection{Data Processing and Training Strategy}
%\todo{add data demographics}
We use pre-processed FDG-PET scans and T1-weighted MRI scans from the Alzheimer's disease neuroimaging initiative (ADNI; \cite{jack2008alzheimer}) database.
Full details about the pre-processing steps can be found at the ADNI website for FDG-PET\footnote{http://adni.loni.usc.edu/methods/pet-analysis} and for MRI\footnote{http://adni.loni.usc.edu/methods/mri-analysis}.
Both scans were additionally processed using SPM\footnote{https://www.fil.ion.ucl.ac.uk/spm/software/spm12} and CAT12 \cite{gaser2016cat}. MRI scans were processed using the standard VBM pipeline in CAT12\footnote{http://www.neuro.uni-jena.de/cat12/CAT12-Manual.pdf}.
% The initial voxel-based processing applies a denoising filter and bias correction to the images, followed by the standard SPM unified segmentation.
% In the second stage, skull-stripping is performed and the brain is parcellated before the final AMAP segmentation step \cite{rajapakse1997statistical}.
% In the last step, the tissue segments are spatially normalized to a common reference space using Geodesic Shooting \cite{ashburner2011diffeomorphic}.
We use the gray matter (GM) tissue area of the brain as an input to the network.
FDG-PET scans were normalized and registered to the MNI152 template \cite{fonov2011unbiased} with 1.5\,mm$^{3}$ voxel size.
We performed min–max scaling to rescale the image intensity values to the range between 0 and 1.
The final image size for both modalities is $113 \times 137 \times 113$. 

% One major source of bias when evaluating deep learning models is data leakage \cite{wen2020convolutional}. Common mistakes highlighted in \cite{wen2020convolutional} include the use of multiple scans per subject, applying data augmentation before data splitting, overfitting on the test set, and differences in the distribution of sex, age, and education across folds. To address these issues, we use a 5-fold cross-validation scheme and adopt the following strategies. Firstly, we only include baseline visits scans for all our splits so that only a single scan per patient is available. Secondly, data augmentation is applied exclusively to the training set and the best performing model is selected based on the accuracy on the validation set. 
% Thirdly, to solve the distribution issue across splits, we assess its balance by computing the propensity score, i.e., the probability of a sample belonging to the training data, based on a logistic regression model comprising age, sex, and education. Next, we compared the percentiles of the propensity score distribution in the training and test data and used the maximum deviation across all percentiles as a measure of imbalance \cite{ho2007matching}. For each of the 5 splits, this process was repeated for 1000 randomly selected partitions and the partition
% with the with minimum imbalance was ultimately the selected split. 
Our dataset comprises 257 patients with AD, 370 healthy controls (CN), and 611 patients with mild cognitive impairment (MCI); see the supplemental Table~\ref{tab:demographic} for additional information.
We split the data into train/validation/test sets with sizes roughly in 65/15/20$\%$ of the full dataset.
We perform cross-validation across 5 splits, based on a data stratification strategy that accounts for sex, age and diagnosis.  We only include baseline visits scans so that only a single scan per patient is available.
We train models for two tasks (i) binary
classification of healthy controls (CN) vs.\ patients with AD,
and (ii) three-way classification of
CN vs.\ MCI vs.\ AD.
All models are trained end-to-end using a cross-entropy loss and data augmentation
during training (up to $8^\circ$ angle rotation and 8\,mm translation in each dimension).
More information about the training setup can be found in the supplemental Table~\ref{tab:train_details}.

\begin{table}[htb]
\scriptsize
\centering
\caption{\label{tab:random_test}%
Overview of the evaluation scheme for correct data, random MRI or random PET.
%Training on correct data and testing on correct, and random PET or MRI.
Numbers are mean balanced accuracy (BACC) and standard deviation across folds.}
\begin{tabular}{lcccc}
\otoprule
              & \ \ Random MRI \ \ & \ \ Random PET \ \ & \ \ BACC 2-Class \   \ & \ \ BACC 3-Class    \\
\midrule
Early Fusion  & \xmark     & \xmark     & $0.885 \pm 0.041$ & $0.573 \pm 0.023$ \\
Early Fusion  & \xmark     & \cmark     & $0.696 \pm 0.026$& $0.414 \pm 0.015$ \\
Early Fusion  & \cmark     & \xmark     & $0.720 \pm 0.015$ & $0.470 \pm 0.028$ \\
\midrule
Middle Fusion & \xmark     & \xmark     & $0.893 \pm 0.036$ & $0.530 \pm 0.034$ \\
Middle Fusion & \xmark     & \cmark     & $0.527 \pm 0.020$ & $0.366 \pm 0.028$ \\
Middle Fusion & \cmark     & \xmark     & $0.890 \pm 0.020$ & $0.528 \pm 0.025$ \\
\midrule
Late Fusion   & \xmark     & \xmark     & $0.896 \pm 0.019$ & $0.577 \pm 0.029$ \\
Late Fusion   & \xmark     & \cmark     & $0.597 \pm 0.029$ & $0.368 \pm 0.027$ \\
Late Fusion   & \cmark     & \xmark     & $0.786 \pm 0.080$ & $0.527 \pm 0.038$ \\
\bottomrule
\end{tabular}
\end{table}

% \begin{table}[htb]
% \centering
% \caption{\label{tab:two_classes}%
% Classification accuracy for CN vs. AD and CN vs. MCI vs. AD
% }
% \setlength{\tabcolsep}{10pt}
% \begin{tabular}{lccc}
% \otoprule
%               & Correct           & Random MRI            & Noise             \\
% \midrule
% \multicolumn{4}{l}{CN vs. AD} \\
% \midrule
% PET only      & $0.905 \pm 0.015$ &      -             &     -              \\
% MRI only      & $0.866 \pm 0.029$ &     -              &       -            \\
% Early   & $0.885 \pm 0.041$ & $0.729 \pm 0.034$ & $0.901 \pm 0.034$ \\
% Middle  & $0.893 \pm 0.036$ &  $0.863 \pm 0.026$ & -                    \\
% Late    & $0.896 \pm 0.019$ & $0.906 \pm 0.022$ & $0.898 \pm 0.028$\\
% \midrule
% \multicolumn{4}{l}{CN vs. MCI vs. AD} \\
% \midrule
% PET only      & $0.541 \pm 0.034$ &      -             &     -              \\
% MRI only      & $0.536 \pm 0.062$ &     -              &       -            \\
% Early   & $0.573 \pm 0.023$ & $0.365 \pm 0.037$ & $0.565 \pm 0.028$ \\
% Middle  & $0.530 \pm 0.034$ &  $0.698 \pm 0.087$      &     -             \\
% Late    & $0.577 \pm 0.029$ & $0.658 \pm 0.015$ & $0.489 \pm 0.051$\\
% \bottomrule
% \end{tabular}
% \end{table}

\section{Results}

\textbf{Testing on Random PET or MRI.} Table \ref{tab:random_test} reports the results for the experiments described in section~\ref{sec:eval_scheme}, for binary and three-way classification.
We observe that when testing on the correct pair of scans, all fusion approaches perform similarly for both tasks with two exceptions: Early Fusion achieves a mean BACC
approximately 0.01 lower for binary classification,
and Middle Fusion a BACC approximately 0.04 lower for three-class classification.
Overall, we  observe a significant drop in performance between these two tasks, which is expected given that MCI is not a true diagnosis, but a syndrome, which makes
it highly heterogeneous, especially
with limited amount of training data.

Interestingly, if we look at the results for the middle and late fusion models when testing on partially random data, we observe a much larger drop in performance when the FDG-PET is randomized; the accuracy is close to random chance.
On the other hand, randomizing the MRI data has much lower impact on the overall performance.
For binary classification the mean BACC
drops around 0.11 for late fusion and
merely 0.003
for middle fusion, which is much lower
than for the random PET experiments:
0.299 and 0.366, respectively.
For early fusion, results for both randomized experiments experience a significant drop compared to using the original data.
% This outcome is expected, since the inputs in both experiments are closely related given that early fusion essentially multiplies the two inputs, which is commutative.
This outcome is expected, since early fusion
results in a single volume where the MRI
acts as a mask to select regions from the FDG-PET.
If the pair of images is from different patients, anatomies are not perfectly aligned
and early fusion will  remove important areas.
Hence, the effect of randomizing the MRI or the FDG-PET leads to a similar loss in information and comparable drop in performance.
% in both random experiments the input to the network is fairly similar (i.e. since the sMRI and the FDG-PET are no longer belonging to the same patient, and therefore not perfectly aligned, the masking operation will occlude areas in the image that might be relevant).

\noindent
\textbf{Training on random MRI:} 
The performance difference between randomizing the FDG-PET data vs.\ the MRI
(see Table~\ref{tab:random_test})
suggests that both modalities do not have the same contribution to the models' final decision.
We decided to further evaluate this hypothesis by defining an additional experiment: during training, the FDG-PET remains associated to a specific patient, but the MRI is exchanged with a random subject. 
Table \ref{tab:two_classes} shows the  results for two- and three-class. 
% classification when training and testing on the correct data (as in Table \ref{tab:random_test}), random MRI, as well as on each single modality. 
Note that results for the original data (Correct) are identical to those in Table~\ref{tab:random_test}.
For binary classification with correct data, 
 middle and late fusion
outperform early fusion by at least 0.08 in mean BACC.
Single modality PET yields the best performance on correct data.
When using a random MRI, the BACC for early fusion decreases, but improves for late fusion, matching the BACC of the single modality PET.
For three-classes with correct data, using PET and MRI data performs similarly with a 0.03 improvement for early and late fusion, while middle fusion decreases in performance by 0.01 compared to using only PET. 
For random MRI, we observe a strong improvement for middle fusion (0.168) and late fusion (0.081),
while the accuracy for early fusion decreases to chance level.
% For comparison, and not reported in the table, the average balanced accuracy for PET only without processing is 0.833. 

% \begin{enumerate}
% \item PET outperforms MRI for single modality network
% \item No increase in acc when combining MRI \& PET for all fusion modalities (early, middle, late)
% \item No significant increase in acc compared to single modality PET.
% \item Random MRI shows a drop for early fusion. Concat shows no drop for random or for training with noise.
% \item Early fusion shows a similar/slightly improved results when trained with noise.
% \item Channel exchange has no implementation with noise. Explain why
% \item For CN vs MC vs AD there's improvement when using fusion compared to uni-modal. However, the best results are achieved for middle fusion with random MRI.
% \end{enumerate}

\begin{table}[htb]
\scriptsize
\centering
\caption{\label{tab:two_classes}%
%Classification accuracy for CN vs. AD and CN vs. MCI vs. AD
Training and testing on correct, and random MRI.
Numbers are mean balanced accuracy and standard deviation across folds.}
\setlength{\tabcolsep}{10pt}
\begin{tabular}{lcccc}
\otoprule
& \multicolumn{2}{c}{CN vs. AD} & \multicolumn{2}{c}{CN vs. MCI vs. AD} \\
\cmidrule(lr){2-3} \cmidrule(lr){4-5}
& Correct           & Random MRI            & Correct & Random MRI             \\
\midrule
PET only      & $0.905 \pm 0.015$ &      ---        & $0.541 \pm 0.034$     &     ---              \\
MRI only      & $0.866 \pm 0.029$ &     ---         & $0.536 \pm 0.062$    &       ---            \\
Early Fusion  & $0.885 \pm 0.041$ & $0.729 \pm 0.034$ & $0.573 \pm 0.023$ & $0.365 \pm 0.037$ \\
Middle Fusion  & $0.893 \pm 0.036$ &  $0.863 \pm 0.026$ & $0.530 \pm 0.034$ &  $0.698 \pm 0.087$   \\              
Late  Fusion   & $0.896 \pm 0.019$ & $0.906 \pm 0.022$ & $0.577 \pm 0.029$ & $0.658 \pm 0.015$ \\
\bottomrule
\end{tabular}
\end{table}

\noindent
\textbf{Post-hoc Explanation via Relevance Maps:}
Relevance maps are a helpful way of assessing the decision-making process of a classification model.
In this work, we use them to quantify how much individual
modalities contribute to the final prediction of the network.
We use Integrated Gradients (IG; \cite{sundararajan2017axiomatic}) because its axiomatic approach allows us to precisely quantify how much the MRI and FDG-PET of a multi-modal CNN contribute to a particular prediction.
Given a patient's images and a baseline,
which is defined by the user (in our case a black volume),
IG computes voxel-wise contributions
by integrating along the path from the baseline input
to the real input.
Since the sum of all voxel-wise IG scores equals
the predicted log-probability, we can summarize
the total contribution of the MRI and FDG-PET by
summing over the IG scores for the respective modality.
Figure~\ref{fig:ig-correct-mri} depicts
the average absolute importance 
across 42 correctly classified AD patients by the late fusion model for CN vs. AD.
This example clearly illustrates that the
PET contributes significantly more to the
overall predictions.
% , thus the network extracts
% most information for prediction from the FDG-PET.
Overall, the PET contributes
1.77 times more to a prediction than the MRI
(sum of |IG| is 33.8 vs.\ 19.1), which confirms
our results from above.

%%%%%%%%%%%%%%%%%%%%%%%%%
% Discussion
%%%%%%%%%%%%%%%%%%%%%%%%%
\section{Discussion}
\begin{figure}[t]
    \centering
    \includegraphics[width=.69\textwidth]{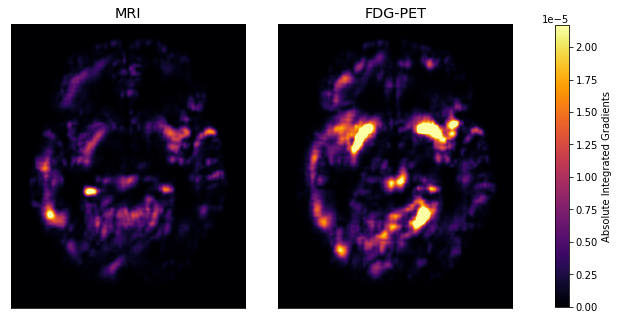}
    \caption{\label{fig:ig-correct-mri}%
    Mean absolute integrated gradients across 42 correctly classified AD patients by the late fusion model.
    Illustrated is an axial slice located at the center
    of the volume.}
\end{figure}

We performed a thorough evaluation of the different methods across 5 splits. 
In our first set of experiments, we observed that when training on correct data but introducing random FDG-PET or MRI data at test time,
both the middle and late networks were
more sensitive to changes of  PET. 
While this is already a strong indicator of the bias of the neural network, our second set of experiments (Table \ref{tab:two_classes}) give us better insights on the reasons behind this phenomenon.
First of all, the AD vs.\ CN classification experiments are consistent across Tables \ref{tab:random_test} and \ref{tab:two_classes},
which proves that the middle and late fusion networks rely mostly on FDG-PET.
These results are supported by the relevance maps in Fig.~\ref{fig:ig-correct-mri}. 

For the three-class experiment,
%with the inclusion of MCI,
the BACC is below 60\%, confirming the difficulty of the task.
MCI subjects are a heterogeneous group that may also suffer from other types of dementia. Therefore, the amount of data required to train a predictive model for this task is much larger than in a two-class setting.
For this challenging task, the usage of random MRIs led to a steep increase in accuracy for middle and late fusion. 
We believe that randomizing the MRI data serves as an augmentation mechanism during training.
Given that in each epoch, the model sees a different pair of FDG-PET and MRI scans, this is likely making the networks more robust to alterations during inference. 

Our results, while being aligned to previous medical findings, are in disagreement with previous literature that favored the fusion of MRI and FDG-PET for AD prediction.
One reason for this difference could be that
randomly exchanging image pairs during training
leads to a larger effective training data size,
which in turn allows the network to be more robust to changes in the data distribution
during inference
% One of the reasons, as previously mentioned, could be that by adding more data to a network through the multi-modal fusion you are just exposing it to more information and therefore making it more robust to changes in the distribution
(similarly observed in Table \ref{tab:two_classes} for the random MRI experiments). 
%Another explanation is %reason could be that by 
Additionally, by increasing the number of branches (e.g. two branches for the late fusion) the amount of trainable parameters is almost doubled, which allows the network to define more complex decision boundaries.
This also makes the networks more prone to overfitting as observed in the three-class experiment when comparing late fusion on correct or random MRI.
Finally, another potential reason is the importance of the PET pre-processing.
For instance, \cite{song2021effective} use a different pre-processing for the image fusion (for which they report high accuracy) and different input for the uni-modal and concatenation networks.
GM is used in image fusion and MNI-MRI for the other combinations.
When we compared the performance between processed and un-processed PET data, we noted a decrease of about 7\% (t-test P=0.01) in balanced accuracy.

\section{Conclusion}
In this work, we rigorously evaluated single- and
multi-modal deep neural networks for AD diagnosis
based on MRI and FDG-PET.
Our results demonstrate that a single-modality network
using FDG-PET performs best for healthy/AD classification.
While this is in contrast with previous work on deep
learning for modality fusion, it does conform with the
established clinical knowledge that FDG-PET better
captures AD-specific patterns of neurodegeneration
than MRI.
We argue that recent work on multi-modal fusion,
while technically sound, are largely disconnected from
the established clinical knowledge about biomarkers in AD.
We argue that future work on multi-modal fusion for AD diagnosis
should take the existing clinical knowledge better into account,
and systematically assess the contribution of individual
modalities following our experimental setup. In the future, we plan to conduct experiments for MCI vs. NC, validate our hypotheses on different datasets and test other classification models.

\paragraph{Acknowledgment}
This research was partially supported by the Bavarian State Ministry of Science and the Arts and coordinated by the bidt, and the Federal Ministry of Education and Research in the call for Computational Life Sciences (DeepMentia, 031L0200A). We gratefully acknowledge the computational resources provided by the Leibniz Supercomputing Centre (www.lrz.de).

\bibliographystyle{splncs03.bst}
\bibliography{biblio.bib}

\begin{thebibliography}{10}
\providecommand{\url}[1]{\texttt{#1}}
\providecommand{\urlprefix}{URL }

\bibitem{Aisen2017}
Aisen, P.S., Cummings, J., Jack, C.R., Morris, J.C., Sperling, R., other: {On
  the path to 2025: understanding the Alzheimer's disease continuum}.
  Alzheimers Res Ther  9(1), ~60 (2017)

\bibitem{basaia2019automated}
Basaia, S., Agosta, F., Wagner, L., Canu, E., Magnani, G., Santangelo, R.,
  et~al.: {Automated classification of Alzheimer's disease and mild cognitive
  impairment using a single MRI and deep neural networks}. NeuroImage Clin  21,
   101645 (2019)

\bibitem{bloudek2011review}
Bloudek, L.M., Spackman, D.E., Blankenburg, M., Sullivan, S.D.: {Review and
  meta-analysis of biomarkers and diagnostic imaging in Alzheimer's disease}. J
  Alzheimers Dis  26(4),  627--645 (2011)

\bibitem{Borson2013}
Borson, S., Frank, L., Bayley, P.J., Boustani, M., Dean, M., Lin, P.J., et~al.:
  Improving dementia care: The role of screening and detection of cognitive
  impairment. Alzheimers Dement  9(2),  151--159 (2013)

\bibitem{Brodersen2010}
Brodersen, K.H., Ong, C.S., Stephan, K.E., Buhmann, J.M.: {The Balanced
  Accuracy and Its Posterior Distribution}. In: 20th International Conference
  on Pattern Recognition. pp. 3121--3124 (2010)

\bibitem{ding2019deep}
Ding, Y., Sohn, J.H., Kawczynski, M.G., Trivedi, H., Harnish, R., et~al.: {A
  deep learning model to predict a diagnosis of Alzheimer disease by using
  18F-FDG PET of the brain}. Radiology  290(2),  456--464 (2019)

\bibitem{ebrahimighahnavieh2020deep}
Ebrahimighahnavieh, M.A., Luo, S., Chiong, R.: {Deep learning to detect
  Alzheimer's disease from neuroimaging: A systematic literature review}.
  Comput Methods Programs Biomed  187,  105242 (2020)

\bibitem{esmaeilzadeh2018end}
Esmaeilzadeh, S., Belivanis, D.I., Pohl, K.M., Adeli, E.: {End-to-end
  Alzheimer’s disease diagnosis and biomarker identification}. In: MLMI. pp.
  337--345 (2018)

\bibitem{farooq2017deep}
Farooq, A., Anwar, S., Awais, M., Rehman, S.: {A deep CNN based multi-class
  classification of Alzheimer's disease using MRI}. In: IST. pp. 1--6 (2017)

\bibitem{feng2019deep}
Feng, C., Elazab, A., Yang, P., Wang, T., Zhou, F., Hu, H., Xiao, X., Lei, B.:
  {Deep learning framework for Alzheimer’s disease diagnosis via 3D-CNN and
  FSBi-LSTM}. IEEE Access  7,  63605--63618 (2019)

\bibitem{fonov2011unbiased}
Fonov, V., Evans, A.C., Botteron, K., Almli, C.R., McKinstry, R.C., Collins,
  D.L., et~al.: Unbiased average age-appropriate atlases for pediatric studies.
  Neuroimage  54(1),  313--327 (2011)

\bibitem{Frisoni2013}
Frisoni, G.B., Bocchetta, M., Chetelat, G., Rabinovici, G.D., de~Leon, M.J.,
  et~al.: {Imaging markers for Alzheimer disease: Which vs how}. Neurology
  81(5),  487--500 (2013)

\bibitem{gaser2016cat}
Gaser, C., Dahnke, R., et~al.: Cat-a computational anatomy toolbox for the
  analysis of structural mri data. Hbm  2016,  336--348 (2016)

\bibitem{hosseini2016alzheimer}
Hosseini-Asl, E., Gimel'farb, G., El-Baz, A.: {Alzheimer's disease diagnostics
  by a deeply supervised adaptable 3D convolutional network}. arXiv preprint
  arXiv:1607.00556  (2016)

\bibitem{huang2019diagnosis}
Huang, Y., Xu, J., Zhou, Y., Tong, T., Zhuang, X., et~al.: {Diagnosis of
  Alzheimer’s disease via multi-modality 3D convolutional neural network}.
  Front Neurosci p. 509 (2019)

\bibitem{Ioffe2015}
Ioffe, S., Szegedy, C.: {Batch Normalization: Accelerating Deep Network
  Training by Reducing Internal Covariate Shift}. In: ICML. pp. 448--456 (2015)

\bibitem{jack2008alzheimer}
Jack~Jr, C.R., Bernstein, M.A., Fox, N.C., Thompson, P., Alexander, G., Harvey,
  D., et~al.: {The Alzheimer's disease neuroimaging initiative (ADNI): MRI
  methods}. J Magn Reson Imaging  27(4),  685--691 (2008)

\bibitem{korolev2017residual}
Korolev, S., Safiullin, A., Belyaev, M., Dodonova, Y.: {Residual and plain
  convolutional neural networks for 3D brain MRI classification}. In: ISBI. pp.
  835--838 (2017)

\bibitem{li2017alzheimer}
Li, F., Cheng, D., Liu, M.: Alzheimer's disease classification based on
  combination of multi-model convolutional networks. In: IST. pp. 1--5 (2017)

\bibitem{liu2018classification}
Liu, M., Cheng, D., Yan, W., et~al.: {Classification of Alzheimer’s disease
  by combination of convolutional and recurrent neural networks using FDG-PET
  images}. Front Neuroinform  12, ~35 (2018)

\bibitem{Livingston2017}
Livingston, G., Sommerlad, A., Orgeta, V., Costafreda, S.G., Huntley, J.,
  et~al.: Dementia prevention, intervention, and care. The Lancet  390(10113),
  2673--2734 (2017)

\bibitem{marcus2014brain}
Marcus, C., Mena, E., Subramaniam, R.M.: {Brain PET in the diagnosis of
  Alzheimer’s disease}. Clin Nucl Med  39(10),  e413 (2014)

\bibitem{payan2015predicting}
Payan, A., Montana, G.: {Predicting Alzheimer's disease: a neuroimaging study
  with 3D convolutional neural networks}. arXiv preprint arXiv:1502.02506
  (2015)

\bibitem{song2021effective}
Song, J., Zheng, J., Li, P., Lu, X., Zhu, G., Shen, P.: {An effective
  multimodal image fusion method using MRI and PET for Alzheimer's disease
  diagnosis}. Front Digit Health  3, ~19 (2021)

\bibitem{sundararajan2017axiomatic}
Sundararajan, M., Taly, A., Yan, Q.: Axiomatic attribution for deep networks.
  In: ICML. pp. 3319--3328 (2017)

\bibitem{Teipel2017}
Teipel, S., Kilimann, I., Thyrian, J.R., Kloppel, S., Hoffmann, W.: Potential
  role of neuroimaging markers for early diagnosis of dementia in primary care.
  Curr Alzheimer Res  15(1),  18--27 (2017)

\bibitem{wang2020deep}
Wang, Y., Huang, W., Sun, F., Xu, T., Rong, Y., Huang, J.: Deep multimodal
  fusion by channel exchanging. NeurIPS  33,  4835--4845 (2020)

\bibitem{yee2020quantifying}
Yee, E., Popuri, K., Beg, M.F., et~al.: {Quantifying brain metabolism from
  FDG-PET images into a probability of Alzheimer's dementia score}. Hum Brain
  Mapp  41(1),  5--16 (2020)

\bibitem{zhang2011multimodal}
Zhang, D., Wang, Y., Zhou, L., Yuan, H., Shen, D., et~al.: {Multimodal
  classification of Alzheimer's disease and mild cognitive impairment}.
  Neuroimage  55(3),  856--867 (2011)

\bibitem{zhou2019effective}
Zhou, T., Thung, K.H., Zhu, X., Shen, D.: Effective feature learning and fusion
  of multimodality data using stage-wise deep neural network for dementia
  diagnosis. Hum Brain Mapp  40(3),  1001--1016 (2019)

\end{thebibliography}

\clearpage
\appendix
\renewcommand\thefigure{S\arabic{figure}}%
\renewcommand\thetable{S\arabic{table}}%
\setcounter{figure}{0}%
\setcounter{table}{0}%
\begin{figure}[t]
    %\centering
    \includegraphics[width=\textwidth]{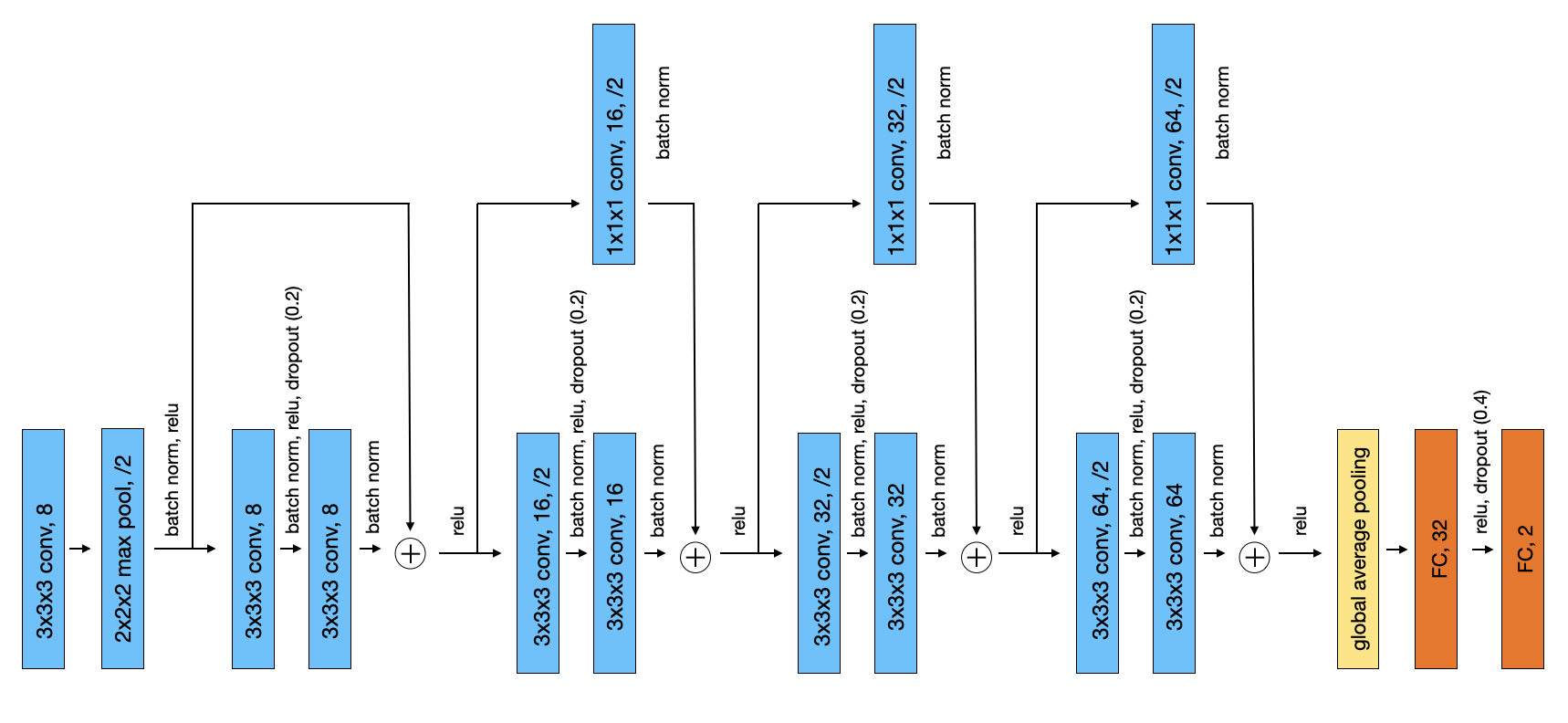}
    \caption{Base architecture used for all the experiments.}
    \label{fig:network_architecture}
\end{figure}

\begin{table}[t]
\centering
\caption{
\label{tab:demographic}%
Dataset statistics.}
% \centering
\begin{tabular}{lcccr}
\otoprule
Subjects                  & Number           & Female/Male     & Age\\
\midrule

CN  & 379 & 193/186 & $ 73.46 \pm 5.93$ \\
MCI  & 611 & 253/358 &  $72.31 \pm 7.30$ \\
AD  & 257 & 104/153 & $74.41 \pm 7.89$ \\
\bottomrule
\end{tabular}
\end{table}

\begin{table}[t]
\centering
\caption{
\label{tab:train_details}%
Overview of the parameters used across all experiments; learning rate was reduced using the ReduceLROnPlateau scheduler from PyTorch. AdamW was used for the optimizer. In middle fusion two additional parameters are used: lambda for the L1 penalty and batch norm threshold for exchanging unimportant channels.}
% \centering
%\setlength{\tabcolsep}{3pt}
\begin{tabular}{lc}
\otoprule
Parameter & Value\\
\midrule
Learning rate & 0.005\\
Epochs & 120\\
Batch size & 16 \\
Weight decay & 0.0001 \\
Batch norm momentum &  0.05\\
Lambda & 0.005 \\
Batch norm threshold  &  0.02\\
\bottomrule
\end{tabular}
\end{table}
\end{document}